\documentclass[aps,prc,twocolumn,superscriptaddress,showpacs]{revtex4}

\usepackage{graphicx}

\begin{document}

\title{Effect of neutron composition and excitation energy on the primary fragment yield distribution in multifragmentation reaction}

\author{D.V.~Shetty}
\affiliation{Cyclotron Institute, Texas A$\&$M University, College Station, TX 77843, USA }
\author{A.S.~Botvina}
\affiliation{Cyclotron Institute, Texas A$\&$M University, College Station, TX 77843, USA }
\affiliation{Institute for Nuclear Research, 117312 Moscow, Russia}
\author{S.J.~Yennello}
\author{G.A.~Souliotis}
\author{E.~Bell}
\author{A.~Keksis}
\affiliation{Cyclotron Institute, Texas A$\&$M University, College Station, TX 77843, USA }
\date{\today}

\begin{abstract}
The isotopic properties of the primary and secondary fragment yield distribution in the multifragmentation of
$^{58}$Fe + $^{58}$Ni and $^{58}$Fe + $^{58}$Fe reactions are studied with respect to the $^{58}$Ni + $^{58}$Ni 
reaction at 30, 40 and 47 MeV/nucleon. The reduced neutron and proton densities from the observed fragment yield 
distribution show primary fragment yield distribution to undergo strongly secondary de-excitations. The effect is small at 
the lowest excitation energy and smallest neutron-to-proton ratio and becomes large at higher excitation energies and 
higher neutron-to-proton ratio. The symmetry energy of the primary fragments deduced from the reduced neutron density 
is significantly lower than that for the normal nuclei at saturation density, indicating that the fragments are highly excited 
and formed at a reduced density. Furthermore, the symmetry energy is also observed to decrease slowly with increasing 
excitation energy. The observed effect is explained using the statistical multifragmentation model.
\end{abstract}

\maketitle

\section{Introduction}
In multifragmentation reactions an excited nuclei expand into vacuum and decay into various light and heavy fragments 
\cite{BOR01,BOW91,AGO96,BEA00}. These fragments when emitted are highly excited and neutron-rich, and subsequently 
undergo de-excitation to cold and stable isotopes. Similar hot nuclei are also produced in the interior of a collapsing star 
and subsequent supernova explosion \cite{BOT04}. The production of these nuclei depends on their internal excitation and 
is sensitive to the symmetry energy part of the binding energy \cite{BOT02}. A slight decrease in the symmetry energy 
coefficient and subsequent de-excitation can significantly alter the elemental abundance and the synthesis of heavy 
elements \cite{BOT04}.  In  multifragmentation reactions the measurement of fragment isotopic yield distribution can provide 
important insight into the symmetry energy and the decay characteristics of these nuclei. 
\par
Experimentally, the determination of fragment isotopic yield distribution is often influenced by the complex de-excitation of 
the primary fragments to their finally observed secondary yield distribution. Theoretical calculations of the yields after secondary 
de-excitations require an accurate accounting of the feeding from the particle unstable state and are subject to uncertainties in 
the levels that can be excited, and the structure effects that govern their decay. 
\par
In this work, we study the primary fragment yield distribution and the associated symmetry energy through  relative 
reduced neutron and proton densities at various excitation energy and isospin (N/Z) of the multifragmenting source. 
It has been shown \cite{XU00} that the relative reduced neutron and proton densities can provide an alternate way of 
studying the isotopic yield distribution, where some of the  uncertainties mentioned above can be avoided by taking the ratios 
of the yield in two different reactions that differ only in their isospin (N/Z) content. Also, the reduced neutron density 
is directly related to the symmetry energy through a scaling relation as discussed in \cite{BOT02,TSAN01,GER04,ONO03}. 
\par
In this work we show that the primary fragment yield distribution is strongly affected by the secondary de-excitations at higher 
excitation energies and isospin (N/Z) of the fragmenting system. The symmetry energy of the primary fragments deduced from 
the scaling parameter is considerably lower than that for the saturation value and decreases with increasing excitation energy. 
The observed effect can be well explained by the statistical multifragmentation model of Botvina {\it {et al.}} \cite{BOT02}. 

\section{Experiment} 
The measurements were carried out at the Cyclotron Institute of Texas A$\&$M University (TAMU) using beams from the K500
Superconducting Cyclotron. Isotopically pure beams of $^{58}$Ni and $^{58}$Fe at 30, 40 and 47 MeV/nucleon were 
bombarded on self-supporting $^{58}$Ni (1.75 mg/cm$^{2}$) and $^{58}$Fe (2.3 mg/cm$^{2}$) targets. Six discrete particle 
telescopes placed inside a scattering chamber at laboratory angles of 10$^{\circ}$, 44$^{\circ}$, 72$^{\circ}$, 
100$^{\circ}$, 128$^{\circ}$ and 148$^{\circ}$ were used to measure the fragments from the reactions. Each telescope consisted 
of a gas ionization chamber (IC) followed by a pair of silicon detectors (Si-Si) and a CsI scintillator detector, providing three 
distinct detector pairs (IC-Si, Si-Si, and Si-CsI) for fragment identification. The ionization chamber was of axial field design and 
was operated with CF$_{4}$ gas at a pressure of 50 Torr. The gaseous medium was 6 cm thick with a typical threshold 
of $\sim$ 0.5 MeV/A for intermediate mass fragments. The silicon detectors had an active area of 5 cm $\times$ 5 cm and were 
each subdivided into four quadrants. The first and second silicon detectors in the stack were 0.14 mm and 1mm thick, respectively.
The dynamic energy range of the silicon pair was  $\sim$ 16 - 50 MeV for $^{4}$He and  $\sim$ 90 - 270 MeV for $^{12}$C. The 
CsI scintillator crystals that followed the silicon detector pair were 2.54 cm in thickness and were read out by photodiodes. Good 
Z identification was achieved for fragments that punched through the IC detector and stopped in the first silicon detector. Fragments 
measured in the Si-Si detector pair also had good isotopic separation. Further details can be found in ref. \cite{SHE03,RAM98}.
\par
The present study is carried out for fragments detected at 44 degrees in the laboratory 
system, which corresponds to the center of mass angle $\approx 90$ degrees. The fragments detected at this angle originate 
predominantly from central events which was further assured by gating on the measured neutron multiplicities. A possible 
admixture of the projectile/target fragmentation processes will nevertheless be accounted for in the following analysis. 
Figure 1 shows the experimentally determined yield distributions for the lithium (left panels) and carbon (right panels) 
isotopes measured in $^{58}$Fe + $^{58}$Fe (triangle symbols), $^{58}$Fe + $^{58}$Ni (circle symbols), $^{58}$Ni + $^{58}$Fe 
(square symbols) and $^{58}$Ni + $^{58}$Ni (star symbols) reactions at the three different beam energies. The data show the highest 
yield for the neutron rich isotopes in the $^{58}$Fe + $^{58}$Fe reaction (triangles), which has the largest neutron-to-proton 
ratio (N/Z),  compared to the $^{58}$Ni + $^{58}$Ni reaction (stars), which has the smallest neutron-to-proton ratio. The yields 
\begin{figure}
\includegraphics[width=0.5\textwidth,height=0.50\textheight]{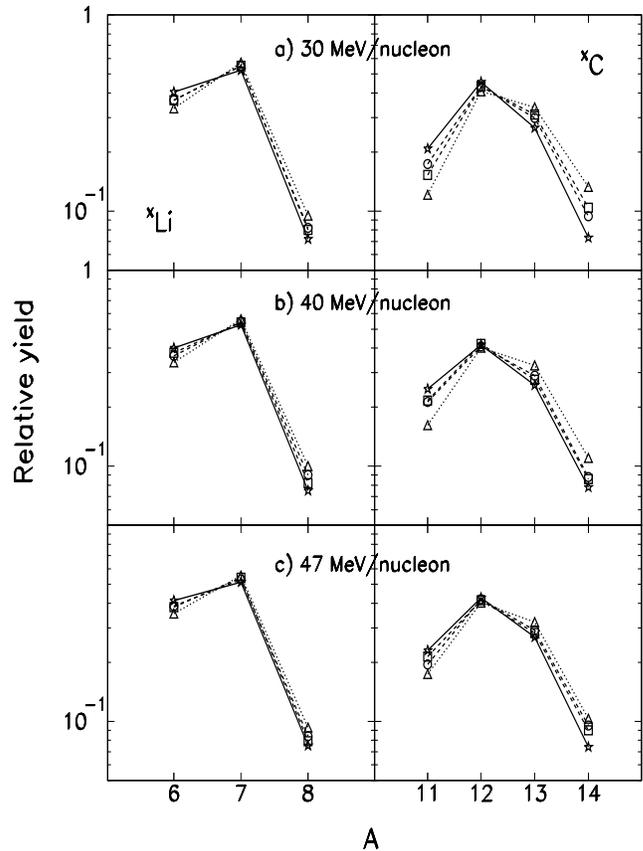}
\caption{Relative yield distribution of the fragments for the lithium (left) and carbon (right) isotopes in $^{58}$Fe + $^{58}$Fe, 
                $^{58}$Fe + $^{58}$Ni, $^{58}$Ni + $^{58}$Fe and $^{58}$Ni + $^{58}$Ni reactions at three different beam energies. 
                The different symbols are for the four reactions as discussed in the text.}
\end{figure}
for the reactions with the intermediate value of the neutron-to-proton ratio, $^{58}$Ni + $^{58}$Fe (squares) and $^{58}$Fe + $^{58}$Ni 
(circles), lie in between. One also observes that the difference in the yield distribution for the four reactions decreases with 
increasing beam energy. As will become evident in the following sections, this is a consequence of the secondary de-excitation 
of the primary fragments which becomes dominant for systems with increasing neutron-to-proton ratio and excitation energy. 
 
\section{The reduced nucleon densities} 
Using the experimentally observed yield distribution, the reduced neutron and proton densities were extracted as follows. In a 
Grand-Canonical approach for the multifragmentation process (see e.g. \cite{BON95,ALB85,RAN81,BOT87}), the fragment 
yield for an isotope with neutron number N, and proton number Z (mass number A = N + Z), is given as

\begin{equation}
 Y(N,Z) \propto V \rho_{n}^{N}\rho_{p}^{Z}Z_{N,Z}(T)A^{3/2}e^{B(N,Z)/T}
\end{equation}

where V is the volume of the system and $\rho_{n}$ $\propto$ e$^{\mu_{n}/T}$ and $\rho_{p}$ $\propto$ e$^{\mu_{p}/T}$, are 
the primary 'free' neutron and proton densities, respectively. The exponents $\mu_{n}$ and $\mu_{p}$ are the neutron and the
proton chemical potentials, and $Z_{N,Z}(T)$ is the intrinsic partition function of the excited fragment. The quantity $B(N,Z)$, is 
the ground state binding energy of the fragment and $T$, is the temperature. In the above formula, the effect of the Coulomb 
interaction on fragment yield is often neglected. By introducing $\rho_{n}$ and $\rho_{p}$, the actual isotope yields are reduced 
to an approximation appropriate for the thermodynamical limit at high excitation energies \cite{BON95}. Hence, they are referred 
to as the  'reduced' densities. 
\par
As discussed in the introduction, taking the ratios of the fragment yield from two different systems which differ only in their 
isospin (N/Z) content reduces the uncertainties in the quantities shown in Eq. 1. For the present analysis all ratio's were 
taken with respect to the least neutron-rich $^{58}$Ni + $^{58}$Ni reaction. The isotopic yield distribution of the fragments in 
terms of relative reduced neutron density is then given as,  

\begin{equation}  
   \frac{ Y(N + k, Z)/Y^{Ni + Ni}(N + k, Z)} { Y(N, Z)/Y^{Ni + Ni}(N, Z)  } = \bigg( \frac{\rho_n}{\rho_{n}^{Ni + Ni}} \bigg )^{k} ,
\end{equation}

where $k$, correspond to the different isotopes used to determine the double ratio, and $Y^{Ni + Ni}$ is the yield for the 
$^{58}$Ni + $^{58}$Ni reaction. A similar expression for the relative reduced proton density from the isotonic yield ratios can 
also be written as

\begin{equation}  
   \frac{ Y(N, Z + k)/Y^{Ni + Ni}(N, Z + k)} { Y(N, Z)/Y^{Ni + Ni}(N, Z)  } = \bigg( \frac{\rho_p}{\rho_{p}^{Ni + Ni}} \bigg )^{k} ,
\end{equation}

Figure 2 (symbols) shows the relative reduced neutron and proton densities as a function of excitation energy (which correspond 
to different beam energies as discussed in the next section) of the fragmenting source for the $^{58}$Fe + $^{58}$Ni and 
$^{58}$Fe + $^{58}$Fe reactions. The various symbols in the figure correspond to the densities obtained using $k$ = 1, $k$ = 2 

\begin{figure}
\includegraphics[width=0.5\textwidth,height=0.50\textheight]{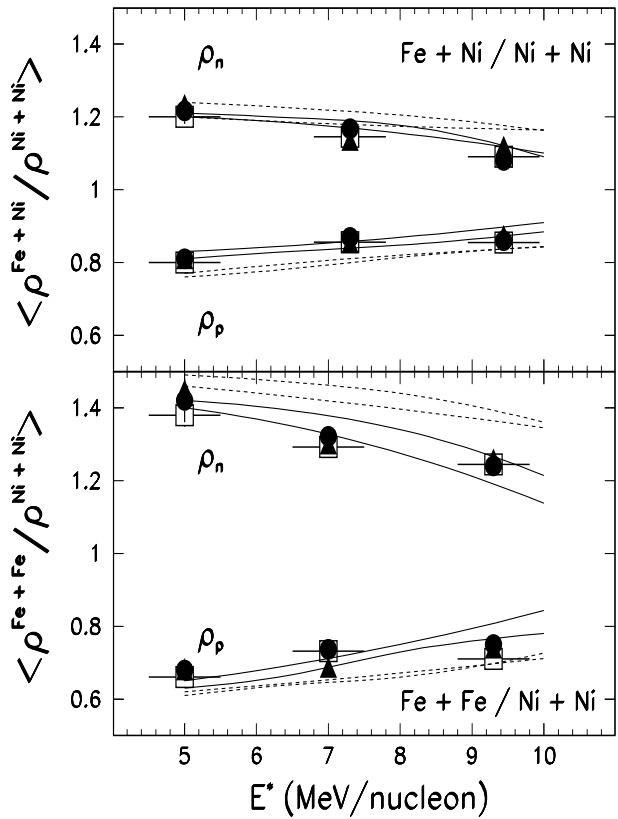}
\caption{Experimentally determined relative reduced neutron and proton densities as a function of excitation energy for the 
         $^{58}$Fe + $^{58}$Ni and $^{58}$Fe + $^{58}$Fe reactions with respect to the 
         $^{58}$Ni + $^{58}$Ni reaction. The different regions bounded by the solid and 
         the dotted lines are the statistical model calculations as explained in the text.} 
\end{figure}

and $k$ = 3, for various isotopes and isotones. The figure shows  a steady decrease in the reduced neutron density and an 
increase in the proton density with increasing excitation energy. The effect is stronger for the $^{58}$Fe + $^{58}$Fe reaction, 
which has the highest isospin (N/Z) content.

\section{Comparison of the reduced nucleon densities with the statistical multifragmentation model calculation}
In order to study the effect of secondary de-excitation at various excitation energies and isospin content (N/Z) of the fragmenting 
system, a comparison of the observed reduced neutron and proton densities with the statistical multifragmentation model (SMM) of 
Botvina {\it { et al.,}} \cite{BOT02} was carried out. The initial parameters for the SMM calculation such as, the mass (A), charge (Z) 
and excitation energy ($E^{*}$) of the thermal source were estimated by simulating the initial stage of the collision dynamics using 
the BNV model calculation \cite{BAR02}. The parameters obtained for the $^{58}$Fe + $^{58}$Ni reaction are, $A_s\approx$ 111, 
$Z_s\approx$ 52 and $E^{*}_s\approx$ 5, 7 and 9.4 MeV/nucleon for the three beam energies studied. The N/Z ratios of the 
sources in all cases were not very different from those of the interacting nuclei. These results were obtained at a time around 
40-50 fm/c after  the projectile fuses with the target nuclei and the quadrupole moment of the nucleon 
coordinates (used for identification of the deformation of the system) nearly goes to zero. Also, the form of single source in the 
calculation oscillates throughout the dynamical evolution and does not undergo any dynamical disintegration, assuring that the 
dynamical flow produced in these reactions are insignificant for the energies studied.
\par
The source parameters were also compared with the experimental data of Hudan {\it {et al.}} \cite{HUD03} where a nearly 
symmetric central collisions of Xe + Sn was studied and an excitation energy of  $\approx$ 5 MeV/nucleon at 32 MeV/nucleon, 
and $\approx$ 7 MeV/nucleon at 50 MeV/nucleon was obtained. It was shown in this work that the SMM calculation describes 
the disassembly of these systems and the excitation energies of the observed primary fragments quite well. A radial flow 
of $\approx$ 2 MeV/nucleon was however identified in this case for the 50 MeV/nucleon beam energy. In the present case, the 
systems are twice as small and stable. Further, it is obvious from various measurements (see e.g. \cite{HON98}) that smaller systems 
accumulate smaller radial flow than larger systems. Consequently, a large part of the energy becomes 
available in the form of thermal excitation. 
\par
To account for the possible uncertainties in the source parameters due to loss of nucleons during pre-equilibrium emission 
and the sensitivity of the final results toward the source size, the calculations were also performed for smaller source sizes of 
$A_s$ = 58 with $Z_s$ = 26, 27, and 28. The freeze-out density in the calculation was assumed to be 1/3 of the normal 
nuclear density and the de-excitation of the hot primary fragments was carried out via evaporation, fission or Fermi 
breakup \cite{BOT87}
\par
Figure 3 shows the mean characteristics of the hot primary fragments in the freeze-out volume calculated from the 
statistical multifragmentation model. Shown are, their relative yields, internal excitation energies and N/Z ratios as a function 
of their mass. At the lowest excitation energy E$^{*}_{s}$ = 6 AMeV, the primary fragments are characterized by an exponential-like 
mass distribution (bottom panel). With increasing excitation of the source the internal excitation energy of the fragments increases 
and the source breaks up into smaller fragments. For sources with different neutron-to-proton ratios (N/Z), the primary fragments 
are also produced with varying N/Z ratios depending on the source excitation energy (top panel). 
\par
The characteristics of the hot primary fragments however, change significantly after the secondary de-excitation as shown in 
Figure 4. Both the mass and the neutron content of the cold secondary fragments decrease considerably lowering the N/Z ratio 

\begin{figure}
\includegraphics[width=0.5\textwidth,height=0.50\textheight]{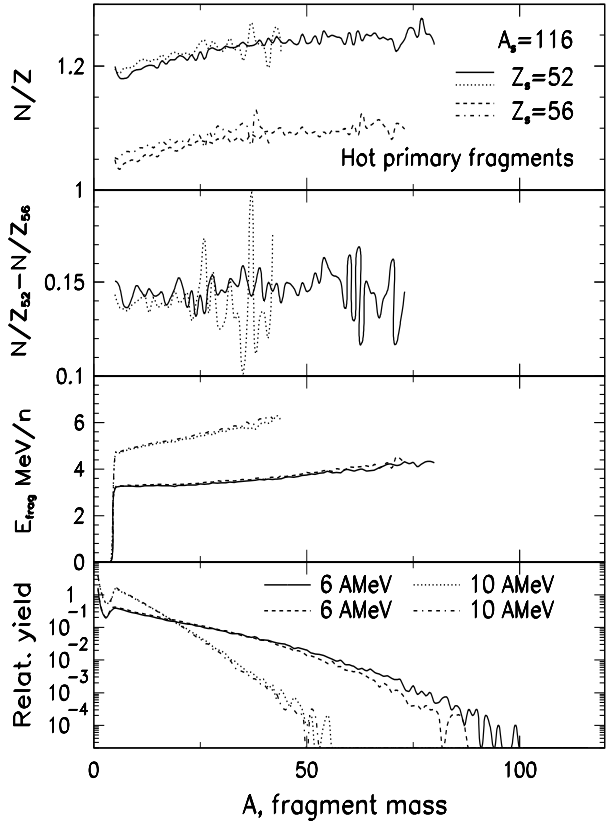}
\caption{Characteristics of the hot primary fragments versus their 
             mass number A, produced in the freeze-out volume during multifragmentation 
                 of the thermal sources, with mass number $A_s$ = 116, and charges $Z_s$ = 52 and 
                 $Z_s$ = 56, at excitation energies of $E^{*}_s$ = 6 and 10 MeV/nucleon (see 
                  notations in the figure). Panels from top to bottom: neutron to proton (N/Z) ratio; 
                 difference of N/Z ratios between neutron rich ($Z_s$ = 52) and neutron poor ($Z_s$ = 56) 
                 sources at different $E^{*}_s$; internal excitation energies; relative mass yields.} 
\end{figure}

\begin{figure}
\includegraphics[width=0.5\textwidth,height=0.50\textheight]{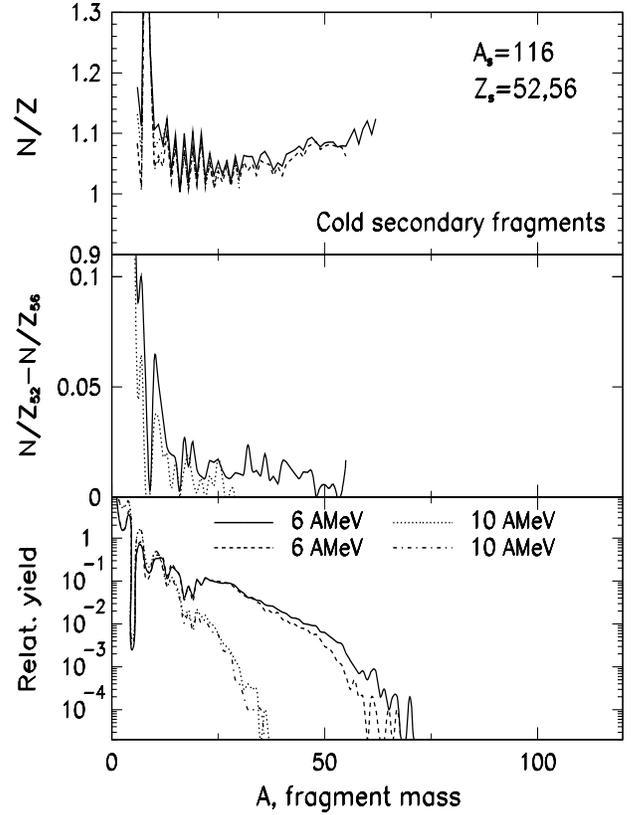}
\caption{Same as in Fig.~3, but for the cold fragments produced after the secondary de-excitation of the primary fragments.}
\end{figure}

due to a higher evaporation probability of neutrons relative to protons. The N/Z of the light fragments vary with Z reflecting their 
shell structure. However, the qualitative difference between the N/Z ratios of the hot fragments (middle panel) produced by the 
sources with different isospin remains the same even after the secondary de-excitation, though it becomes smaller. Moreover, 
one can see the solid and the dotted lines in the middle panel to be more distinguished than in the Figure 3. The average difference 
between the cases with the two excitations increases from 0.0051 to 0.0168 for the fragments A = 6 - 20. This means that the 
difference in the N/Z of the fragments produced between the neutron rich and the neutron poor sources decreases more rapidly 
for high excitation energies as a result of more intensive secondary de-excitation. 
\par
Figure 2 presents the comparison between the experimentally (symbols) determined relative reduced neutron and proton densities 
and the calculations from the statistical multifragmentation model (lines). The regions enclosed  between the solid lines correspond 
to the SMM calculation obtained from the secondary fragment yield distribution, and those between the dotted lines correspond 
to the densities obtained from the primary fragment yield distribution. The width of the two regions represents an upper limit to the 
sensitivity of the calculation to the assumed source size. The error bars for the excitation energies corresponds to  two different 
equations of state used in the BNV calculation, namely, the asy-stiff and the asy-soft equation of state. It is observed that the 
statistical multifragmentation model calculation from the secondary fragment yield distribution explains the observed data quite 
well. The calculated densities from the primary and the secondary fragment distribution are not very different for 
the $^{58}$Fe + $^{58}$Ni reaction. But those for the $^{58}$Fe + $^{58}$Fe reaction, which has a higher neutron-to-proton (N/Z) 
content, are significantly different. The difference is small at lower excitation energy and becomes larger with increasing excitation 
energy. The comparison shows that the primary fragment yield distribution for systems with large neutron-to-proton (N/Z) ratio and 
initial excitation energy are strongly affected by the secondary de-excitation process. With increasing excitation energy and 
neutron-to-proton ratio of the fragmenting system, the primary fragments are highly neutron rich and produced with large internal 
excitation energy which quickly undergoes secondary de-excitation leading to a decrease in the width of the yield distribution as 
observed in Figure 1. Such an effect has also been observed for fragments produced in a dynamical stochastic mean field 
calculation \cite{BAR02,LIU04}.

\section{The Reduced neutron density and the symmetry energy}
In the following, we extract the symmetry energy of the primary fragments from the reduced neutron density and study their dependence 
on the excitation energy of the fragmenting system. From Eq. 1 the ratio of the fragment yield from two different reactions can be written 
as

\begin {equation}
  \frac{ Y(N, Z)}{Y^{Ni + Ni}(N, Z)}  = C \bigg( \frac{\rho_n}{\rho_{n}^{Ni + Ni}} \bigg )^{N}  \bigg( \frac{\rho_p}{\rho_{p}^{Ni + Ni}} \bigg )^{Z} = C \hat{\rho}_{n}^{N} \hat{\rho}_{p}^{Z}
\end{equation}

Where C is the overall normalization factor and $\hat{\rho}_{n}$, $\hat{\rho_{p}}$ are the relative reduced neutron and proton 
densities. The relative reduced neutron and proton densities are related to the isoscaling parameters 
$\alpha$ = $\Delta$$\mu_{n}/T$ and $\beta$ = $\Delta$$\mu_{p}/T$ through a relationship given by 
$e^{\alpha}$ = $\hat{\rho}_{n}$ and $e^{\beta}$ = $\hat{\rho}_{p}$. The quantities $\Delta$$\mu_{n}$ and $\Delta$$\mu_{p}$ are 
the difference in the neutron and the proton chemical potential between the two reactions. The parameter $\alpha$ is related to the 
symmetry energy part of the fragment binding energy C$_{sym}$, and given as (see \cite{BOT02}),

\begin{equation}
       \alpha T = 4 C_{sym}\bigg (\frac{Z_{1}^{2}}{A_{1}^{2}} - \frac{Z_{2}^{2}}{A_{2}^{2}} \bigg )
\end{equation}

where $Z_{1}$, $A_{1}$ and $Z_{2}$, $A_{2}$ are the charge and the mass numbers of the fragmenting systems. Figure 5 (top panel) 
shows the experimentally (symbols) determined isoscaling parameter $\alpha$, from the reduced neutron density as a function of 
excitation energy of the fragmenting system. Also shown in the figure are the calculated values of the scaling parameter from the 
primary (dashed curve) and the secondary (solid curve) fragment yield distribution using the statistical multifragmentation model. 
Within the uncertainties in the input parameters of the SMM calculation, the decrease in the $\alpha$ values with increasing 
excitation energy is well explained by the calculation once 
\begin{figure}
\includegraphics[width=0.5\textwidth,height=0.50\textheight]{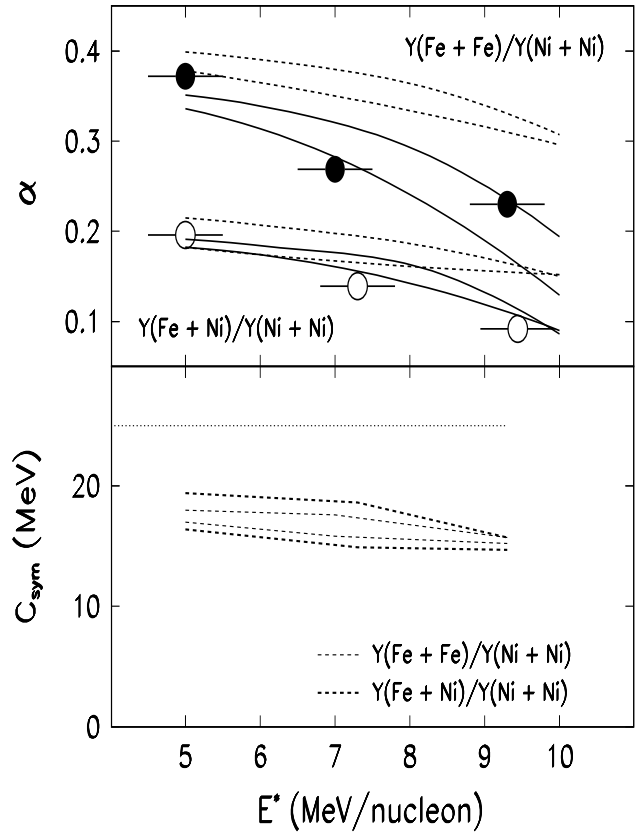}
\caption{Isoscaling parameter (top panel) and symmetry energy (bottom panel) as a function of  excitation energy of the fragmenting source for the 
               $^{58}$Fe + $^{58}$Ni and $^{58}$Fe + $^{58}$Fe reactions. (Top panel) The dashed and the solid curves are the calculated values of 
               $\alpha$ from the primary and secondary fragment yield distributions using SMM, respectively. (Bottom panel) Symmetry 
               energy calculated using the primary fragment yield distribution.} 
\end{figure}
the secondary de-excitation effect in the primary distribution is taken into account. Also, the calculated 
values from the secondary fragment distribution are not very different from those calculated from the primary fragment 
distributions for the lower excitation energies. However, with increasing excitation energy and isospin (N/Z) of the 
system they differ significantly. More specifically, the scaling parameter $\alpha$ for the neutron-rich primary 
fragments produced at  excitation energies greater than 5 - 6 MeV/nucleon are strongly affected by the secondary 
decay effect.   
\par
Having demonstrated the fact that the observed scaling parameter is well reproduced by the statistical multifragmentation model
and that a high degree of thermalization is expected during the fragment formation, we have attempted to obtain an estimate of the 
symmetry energy C$_{sym}$ of the primary fragments using equation 5.  The temperature for the estimate was determined using the 
double isotope ratio of the fragment yield \cite{ALB85}. Since the determination of the temperature from the fragment yield is also 
affected by the secondary de-excitation, a maximum correction of 70$\%$ was applied to be consistent with the previous determination 
of temperature for similar reactions (see \cite{NAT02}). This correction results in a temperature value of 6 - 7 MeV over the energy 
range studied. The difference $(Z/A)_{1}^{2} - (Z/A)_{2}^{2}$ of the fragmenting system was estimated at t = 50 fm/c using the 
dynamical model calculations \cite{ONO03,BAR02,BOTV95} and was about a 3$\%$ lower than the difference in the initial Z/A of 
the systems. Within these uncertainties, one obtains C$_{sym}$ between 15 - 20 MeV over the excitation energy range studied. The 
estimated values from both the systems, $^{58}$Fe + $^{58}$Fe and $^{58}$Fe + $^{58}$Ni, are comparable to each other and 
shown by the thin and thick dashed lines in Fig. 5 (bottom panel). These observed values are significantly lower than the standard 
value of 25 MeV (dotted line) often assumed for a stable and unexcited nuclei. Furthermore, the symmetry energy is also observed to 
decrease slowly with increasing excitation energy. It should be mentioned that the estimated value of the symmetry energy is 
sensitive to the corrections assumed for the $(Z/A)_{1}^{2} - (Z/A)_{2}^{2}$ in Eq. 5. A lower value of the 
$(Z/A)_{1}^{2} - (Z/A)_{2}^{2}$ could raise the value of symmetry energy. However, from the present observation it would require 
a correction of more than 70$\%$ at the highest excitation energy to reproduce the standard value of 25 MeV. Such an increase 
in not supported by any dynamical calculations and appears unlikely. It thus appears impossible to reproduce the standard value 
of the symmetry energy by any means. The data along with the calculations thereby suggests that the symmetry energy is below 
20 MeV and well below the standard value. The reduced value of the symmetry energy shows that the primary fragments are not 
only excited and neutron rich but also expand to a reduced density when formed. For higher excitation energies the fragments 
appear to expand to even lower symmetry energies. A self consistent check of the SMM calculation shows strong variations in the 
properties of the primary fragments when C$_{sym}$ is varied in the range 15 - 25 MeV. Recently \cite{BOT04}, it has been shown 
that neutron rich hot nuclei can be produced in stellar matter between the protoneutron star and the shock front in type II supernova 
explosion. In particular, it has been shown that a slight decrease in the symmetry energy coefficient can shift the mass distributions 
to higher masses. This property of hot nuclei could thus be interesting to investigate for understanding the relative abundance of 
elements in the core collapse supernovae explosion.

\section{Conclusion} 
In conclusion, we have studied the isotopic properties of the primary and secondary fragment yield distribution using the 
reduced nucleon densities in multfragmentation of $^{58}$Ni + $^{58}$Ni, $^{58}$Fe + $^{58}$Ni and $^{58}$Fe + $^{58}$Fe 
reactions at 30, 40 and 47 MeV/nucleon. It is shown that the primary fragment yield distribution for system with large excitation 
energy and neutron-to-proton ratio are significantly affected by 
the secondary de-excitation effects. The effect is small at lower excitation energy and smaller neutron-to-proton ratio 
and becomes large with increasing excitation and neutron-to-proton ratio. The symmetry energy for the fragments 
deduced from the scaling parameter after correcting for the secondary decay effect is considerably lower than 
the standard value at saturation density. Furthermore, the symmetry energy decreases with increasing excitation energy 
indicating that the fragment nuclei are hot and expand to a reduced density when formed. The present observation could 
be important in the understanding of the mechanism of the core collapse supernovae explosions and the relative abundance 
of the produced elements.

\section{Acknowledgment}
The authors wish to thank the staff of the Texas A$\&$M Cyclotron facility for the excellent beam quality. This work was 
supported in part by the Robert A. Welch Foundation through grant No. A-1266, and the Department of Energy through grant 
No. DE-FG03-93ER40773. One of the authors (ASB) thanks Cyclotron Institute TAMU for hospitality and support.


\begin{thebibliography}{}
\bibitem{BOR01} B.~Borderie et al., Phys. Rev. Lett. 86, (2001) 3252. 
\bibitem{BOW91} D.~R.~Bowman et al., Phys. Rev. Lett. 67, (1991) 1527. 
\bibitem{AGO96} M.~D'Agostino et al., Phys. Lett. B 371, (1996) 175. 
\bibitem{BEA00} L.~Beaulieu et al., Phys. Rev. Lett. 84, (2000) 5971. 
\bibitem{BOT04} A.S.~Botvina and I.N.~Mishustin, Phys. Lett. B 584, (2004) 233.
\bibitem{BOT02} A.S.~Botvina, O.V.~Lozhkin, and W.~Trautmann, Phys. Rev. C 65, (2002) 044610. 
\bibitem{XU00} H.~Xu et al., Phys. Rev. Lett. 85, (2000) 716. 
\bibitem{TSAN01} M.B.~Tsang et al., Phys. Rev. C 64, (2001) 054615. 
\bibitem{GER04} E.~Geraci et al., Nucl. Phys. A 732, (2001) 173. 
\bibitem{ONO03} A.~Ono et al., Phys. Rev. C 68, (2003) 051601.
\bibitem{RAM98} E.~Ramakrishnan et al., Phys. Rev. C 57, (1998) 1803.
\bibitem{SHE03} D.V.~Shetty et al., Phys. Rev. C 68, (2003) 021602.
\bibitem{BON95} J.P.~Bondorf et al., Phys. Rep. 257, (1995) 133.
\bibitem{ALB85} S.~Albergo et al., Nuovo Cimento A 89, (1985) 1.
\bibitem{RAN81} J.~Randrup and S.~Koonin, Nucl. Phys. A 356, (1981) 223.
\bibitem{BOT87} A.S.~Botvina et al., Nucl. Phys. A 475, (1987) 663.
\bibitem{BAR02} V.~Baran, M.~Colonna, M.Di~Toro, V.~Greco, M.~Zielinska~Pfabe, and H.H.~Wolter, Nucl. Phys. A 703, (2002) 603.
\bibitem{HUD03} S.~Hudan et al., Phys. Rev. C 67, (2003) 064613. 
\bibitem{HON98} B.~Hong et al., Phys. Rev. C 57, (1998) 244. 
\bibitem{LIU04} T.X.~Liu et al., Phys. Rev. C 69, (2004) 014603
\bibitem{NAT02} J.B.~Natowitz et al., Phys. Rev. C 65, (2002) 034618
\bibitem{BOTV95} A.S.~Botvina , A.B.~Larionov, and I.N.~Mishustin, Phys. of Atom. Nuclei B 58, (1995) 1703.
\end{thebibliography}
\end{document}